\documentclass[preprint,12pt,a4paper]{elsarticle}

\usepackage{amssymb}
\PassOptionsToPackage{hyphens}{url}
\usepackage{hyperref}
\usepackage{amsthm,amsmath}
\usepackage{graphicx}
\usepackage{xspace}
\usepackage{listings}
\usepackage{multirow}
\usepackage{url}

\journal{arXiv}

\newcommand{\name}{{\scshape phyloDB}\xspace}

\begin{document}
\sloppy
\renewcommand{\labelenumii}{\arabic{enumi}.\arabic{enumii}}

\begin{frontmatter}
\title{phyloDB: A framework for large-scale phylogenetic analysis}

\author[label1]{Bruno Lourenço}
\author[label2,label3]{Cátia Vaz}
\author[label3]{Miguel E. Coimbra}
\author[label1,label3]{Alexandre P. Francisco}
\address[label1]{Instituto Superior Técnico, Universidade de Lisboa}
\address[label2]{Instituto Superior de Engenharia de Lisboa, Instituto Politécnico de Lisboa}
\address[label3]{INESC-ID, Lisboa}

\begin{abstract}
\name  is a modular and extensible framework for large-scale phylogenetic analyses, which are  essential for understanding epidemics evolution. It relies on the Neo4j graph database for data storage and processing, providing a schema and an API for representing and querying phylogenetic data. Custom algorithms are also supported, allowing to perform heavy computations directly over the data, and to store results in the database. Multiple 
computation results are stored as multilayer networks, promoting and facilitating comparative analyses, as well as avoiding unnecessary ab initio computations.
The experimental evaluation results showcase that \name is efficient and scalable with respect to both API operations and algorithms execution.
\end{abstract}

\end{frontmatter}

\begin{table}[!ht]
{\small
\begin{tabular}{|l|p{5.6cm}|p{6.0cm}|}
\hline
\textbf{Nr.} & \textbf{Code metadata description} & \textbf{Please fill in this column} \\
\hline
C1 & Current code version & v1.2.0 \\
\hline
C2 & Permanent link to code/repository used for this code version & \url{https://github.com/phyloviz/phyloDB} \\
\hline
C3  & Permanent link to Reproducible Capsule & n/a\\
\hline
C4 & Legal Code License   &  GNU GPL v3 \\
\hline
C5 & Code versioning system used & git \\
\hline
C6 & Software code languages, tools, and services used & Java, Neo4j \\
\hline
C7 & Compilation requirements, operating environments \& dependencies & JDK 11, Spring Boot, jackson-core, Neo4j, Gradle, Apache Maven\\
\hline
C8 & If available Link to developer documentation/manual &  \url{https://github.com/phyloviz/phyloDB/wiki} \\
\hline
C9 & Support email for questions & cvaz@cc.isel.ipl.pt; aplf@inesc-id.pt \\
\hline
\end{tabular}
}
\caption{Code metadata (mandatory).}
\label{codeMetadata} 
\end{table}

\section{Motivation and significance}
\label{sec:background}
Modern biomedical research and engineering has seen a remarkable increase in the production and computational analysis of large datasets, leading to an urgent need of efficient and scalable tools for data integration, processing and visualization, as well as the need of sharing standardized analytical techniques. 
One important field of biomedical research is understanding the evolution of pathogens in order to determine their origin, evolution and resistance. 
For instance, phylogenetic analyses have been essential to determine transmission chains and, even more so, to understand the origin of the migration of SARS-CoV-2 to humans~\cite{rothan2020epidemiology}.

When performing large-scale phylogenetic analyses of microbial population genetics, it is often needed to sequence and type isolates, and afterwards to apply a set of phylogenetic inference methods~\cite{artc:10.1093/bib/bbaa147} to produce a diagrammatic hypothesis about the evolutionary history. 
The computation and analysis of microbial population genetics often produces phylogenetic trees or networks~\cite{artc:trees}. The appearance of NGS technologies further increased this challenge due to the substantial growth in the amount of genomic data that can be used to characterize a population.

The integration of the results obtained from inference algorithms with epidemiological data (also called isolate ancillary data) and simultaneous analysis is still limited by visualization and processing techniques. 
Although there are some tools for visualizing and analyzing such data, allowing the integration of epidemiological data, such as SplitsTree4~\cite{huson2006application}, Phylogeny.fr~\cite{dereeper2008phylogeny}, PHYLOViZ~\cite{artc:phyloviz2.0,artc:phylovizonline} and GrapeTree~\cite{artc:grapetree}, they do not scale for large data analysis and visualization.
Most of them run inference and/or visualization optimization tasks on the client side, requiring data to be transferred from existing databases in order to be analysed.
This is often unfeasible for large amounts of data and, moreover, results and optimizations are not stored for reuse.

Graph databases, such as Neo4j~\cite{artc:neo4j}, address the problem of leveraging complex and dynamic relationships in highly connected data such as the phylogenetic trees or networks.  When the storage is native, such as in Neo4j, their performance tends to remain relatively constant, in contrast to the traditional relational databases, even as the dataset grows. The reason for this performance in graph databases is related to the fact that queries are local to a portion of the graph, and thus execution time for each query is proportional to the size of that part. 
In terms of flexibility, a graph database supports the addition of new nodes, labels and relationships to an existing structure, without jeopardizing existing queries and application functionality. These features are very important for phylogenetic context, since a native graph storage is optimized and designed for storing and managing graphs, and with a native processing engine, query times are proportional to the amount of the graph data searched (independent of the total size of the graph). 

With respect to semantic extensibility,  Neo4j, for instance, provide mechanisms that
allow the extension of the database semantics, by implementing custom functions or procedures. 
This is crucial in large scale phylogenetic analysis, enabling the avoidance of data transfer and loading for phylogenetic analyses, and allowing for more rich algorithms to be implemented and efficiently deployed over the data.

Although graph-oriented databases can be of much help for biomedical data~\cite{sakr2011graph, yoon2017use}, as far as we know there is no solution relying on these technologies to address large-scale phylogenetic analysis challenges.

Thus, the contribution of this work is \name, a modular framework for large-scale phylogenetic analyses, exploiting the Neo4j graph-oriented database to allow for the management of the phylogenetic data, without needing to load it into client computers, or secondary servers, for further exploration. 
\name has a graph data model that allows the representation of phylogenetic data, inferred phylogenetic trees and networks, as well as related ancillary data. 
It efficiently supports queries over such data, and allows the deployment of algorithms for inferring/detecting patterns and for pre-computing and optimizing phylogenetic visualizations. 
By storing the results of algorithms with the data, it also enables the reuse and comparison of results.

\section{Software description}
\label{sec:approach}
\name  is a framework that allows to store and manage data resulting from phylogenetic analyses in a Neo4j graph database. This framework supports the execution of inference and visualization algorithms directly over the stored data offloading heavy computations to the server side. This is possible by supporting algorithms as Neo4j plugins, i.e., user-defined procedures in Neo4j. A user-defined procedure is a mechanism that allows to extend Neo4j by writing custom code, which can be invoked directly from its query language Cypher. \name
provides also a secured API for phylogenetic data management.

\subsection{Software architecture}
\begin{figure}[!t]
 \centering
 \includegraphics[scale=0.66]{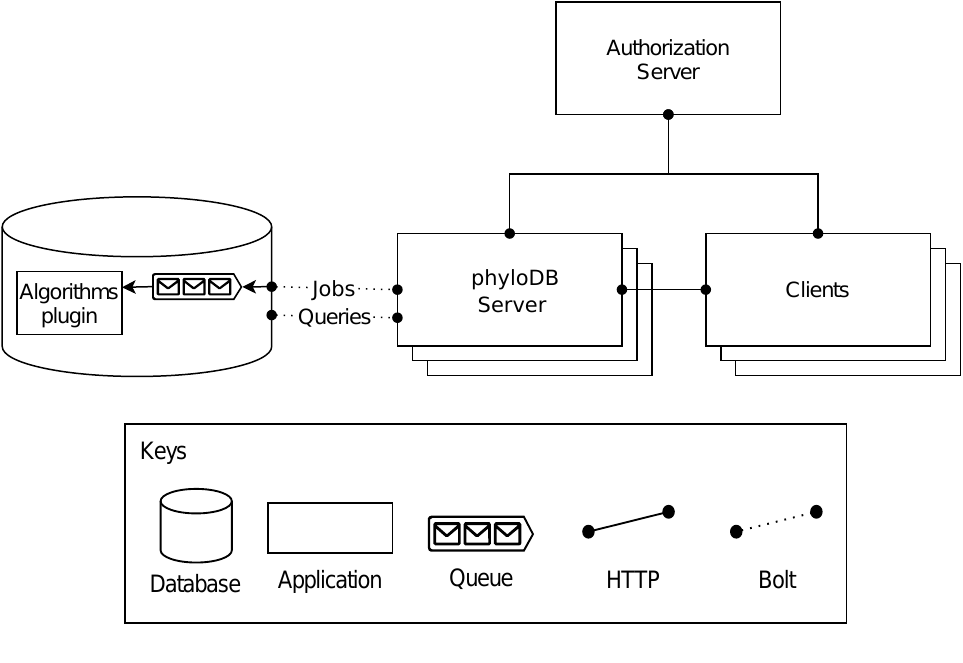}
 \caption{Client-Server architectural view.}
 \label{fig:clientserver}
\end{figure}

This framework consists of several components as depicted in Figure~\ref{fig:clientserver}.
The server component provides a Spring~\cite{web:spring} web application programming interface (API) to perform several operations over the data stored in the database, namely data access and loading, the execution of algorithms and retrieval of results.
It can scale horizontally by adding more instances.
This component interacts with the database component, namely with the Neo4j graph database, for data storage, data management and for queuing the execution of algorithms deployed as plugins.
These algorithms can also read data from the graph database, and write back computed results. 
The authorization component manages the user information and validates the operations available to the authenticated user.
Although we can use any authentication provider, the present implementation relies on Google Identity Provider~\cite{web:googleidp} and on a Keycloak Provider~\cite{thorgersen_2023_keycloak} for authentication.

\subsubsection{Data storage}\label{sec:approach:sec:data_storage}

\begin{figure}[!t]
 \centering
 \includegraphics[scale=0.43]{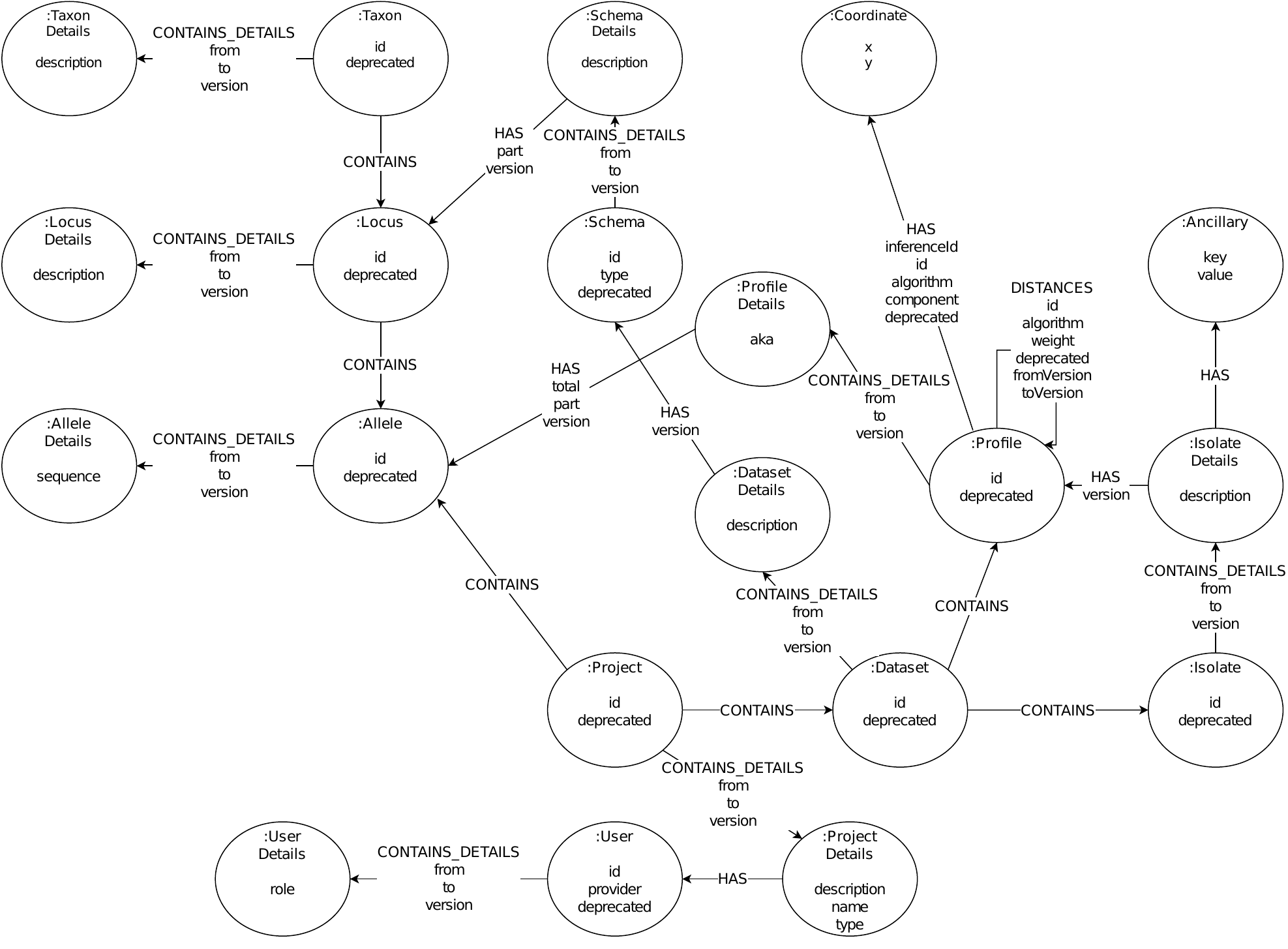}
 \caption{Data model represented by a set of nodes and relationships to compose a graph data model, which is the format used by Neo4j.}
 \label{fig:datamodel}
\end{figure}

We rely on a graph data model built from a subset of the concepts and properties defined in the TypOn ontology~\cite{artc:relations}, allowing the representation of the main entities in phylogenetic analyses as well as their relationships (see Figure~\ref{fig:datamodel}).
Other entities such as users and projects were introduced to support user and project management, including authorization and data versioning.
Moreover, we also have introduced entities relevant to store results from inference and visualization algorithms. For instance, considering distance-based inference methods, one common step of these methods is the calculation of a distance matrix that reflects the pairwise genetic distance between nodes of the phylogenetic tree or network. 
In the phylogenetic context, nodes represents profiles, i.e, characterisations of the genetic sequences of isolate data.
Given a distance matrix, an inference algorithm is executed and links among nodes can be created. Since links are labeled accordingly to the id of the performed inference, different inference results can be registered, and we get a multilayer network among nodes.

Visualizations of inferred patterns represented by links are optimized and rendered through methods such as the Radial Layout~\cite{book:radial} or GrapeTree~\cite{artc:grapetree}.
The optimization process may take considerable time, and usually it can be shared and reused.
This motivates the interest in separating the computation of the optimal coordinates for each node in the visualization from the rendering process.
\name supports the execution of visualization optimizations over the data on the server side, and it stores obtained coordinates in the database for each different layout algorithm execution. 

The data model also supports versioning and soft deletion.
Given the data dependencies, the importance of keeping track of changes and for the sake of reproducible results, we avoid deleting information in the database.
In this case, by considering a versioning and a soft-delete strategy, information removal is possible while keeping previous results valid for the underlying version of the data. 
The versioning strategy to achieve this behavior is to separate each object from its state, link them through a relationship with the respective version number, and capture changes by having different state nodes~\cite{web:neo4jversioning}. The data model is detailed in the \name repository \url{https://github.com/phyloviz/phyloDB/wiki}.

\subsection{Software functionalities}
The main functionalities of \name are (1) a secured API to perform data management operations over the data stored in the graph database, such as data access, querying, loading, and the execution of inference and visualization optimization tasks, storing their results; (2) data versioning; and (3) a plugin interface for adding new inference and visualization algorithms.
  
\subsubsection{Web API}\label{sec:approach:sec:model}
The API was designed not only to allow operations over the data stored in the graph database, but also to support algorithms execution and to provide an easy integration of the framework with other applications such as front-end applications or workflow systems.

The API provided by \name was implemented based on three layers, namely \verb|Controllers|, \verb|Services| and \verb|Repositories|.
When a request is received by the API, it is passed through the \verb|Controllers| layer. 
This layer contains the controllers that parse the received input, execute the respective service, and retrieve the response containing the respective status code and the formatted content. 
The \verb|Services| layer performs the business logic and depends on operations over repositories. 
The \verb|Repositories| layer makes available the operations to interact with the database. 
Apart from these layers, there is a validation logic that verifies the request authenticity and the user permissions before the request is processed. 
The API relies on the REST architecture, and on the HTTP semantics. The set of endpoints that were defined are described in the \name repository \url{https://github.com/phyloviz/phyloDB/wiki}.

The authentication is based on bearer token authentication~\cite{web:bearer}. 
This type of authentication is based on tokens that are acquired after an authentication process with an identity provider.

We note that several aspects were considered in the implementation of repositories, such as the use of an object graph mapper (OGM), pagination and parameterized queries, among others.
The use of the functionalities related to an OGM may add some overhead~\cite{web:cypherfast}.
Hence, it was decided that each query should be implemented from scratch to increase the performance of the data access operations. 
The implementation of each repository method contains then the respective \texttt{Cy\-ph\-er} query that it should perform.
Moreover, the use of pagination in these queries was adopted because this framework is intended to handle large quantities of data. 
Parameterized queries are also used so for better performance, as Neo4j can cache the query plans and reuse them in subsequent executions, increasing the speed of the next query. 
This also enables protection against injection attacks, since parameters are never allowed to be interpreted as part of the query and have no means of escaping out of being anything other than a value of some sort~\cite{web:cypherfast, web:cypherparameterized}. 

\subsubsection{Algorithms as plugins}\label{sec:approach:sec:plugin}

The plugin functionality relies on the user-defined procedures from Neo4j and on the APOC library~\cite{web:neo4japoc}.
A user-defined procedure is a mechanism that extends Neo4j through custom code.
These procedures can take arguments, perform operations on the database and return results. They can be also invoked directly from \texttt{Cy\-ph\-er}. 

In our work plugins intend to extend Neo4j to support inference and visualization algorithms, which shall be available as procedures. 
The inference algorithm procedures are executed over the profiles of a dataset, while visualization algorithm procedures are executed over the results of inference algorithms. 

The structure of an algorithm plugin is also based on three layers, namely (1) the \texttt{Procedures} layer that provides the fundamental operations needed for user-defined algorithms; (2) the \texttt{Services} layer responsible for reading input data from the database, executing algorithms, and storing results back on the database; (3) and the \texttt{Repositories} layer that provides methods to interact with the database.
Once implemented as described below, an algorithm can be compiled and packaged in a jar file, and deployed in the Neo4j \texttt{plugins} directory together with other plugins and the APOC library.

\section{Illustrative examples}
We illustrate in this section the addition of a new algorithm. Further details and examples are provided in the project documentation, including the API documentation and use cases.

Let us add goeBURST~\cite{artc:eburst} algorithm as an example of an inference algorithm. In this context, it must be taken the following steps: (1) create a new procedure; (2) create a new service; (3) create a new repository; (4) compile and package in a jar file. See the project repository for the full example.

\subsection{New Procedure}
Different procedures can be added by creating sub-types of the procedure type.
As depicted in Listing~\ref{lst:Procedure}, \texttt{Procedure} is defined as an abstract type that provides the common base for all procedures.
\begin{lstlisting} [language=Java, caption=Procedure., label=lst:Procedure, frame=single, basicstyle=\scriptsize] 
public abstract class Procedure {
  @Context
  public GraphDatabaseService database;
  @Context
  public Log log;
}
\end{lstlisting}

In Listing~\ref{lst:InferenceProcedure}, for inference algorithms, a new sub-type is defined. Methods declared within this new type must be annotated with the \texttt{@Procedure} annotation to be executed as a standard procedure. 
Also, with this annotation, the designation of the new standard procedure is defined. 
In the example of Listing~\ref{lst:InferenceProcedure}, to execute the new standard procedure \texttt{goeBURST}, the designation to use is \texttt{algorithms.inference.goeburst}.
\begin{lstlisting}[caption=InferenceProcedure., label=lst:InferenceProcedure, frame=single, basicstyle=\scriptsize] 
public class InferenceProcedure extends algorithm.utils.Procedure {
  /*...*/
  @Procedure(value = "algorithms.inference.goeburst", 
             mode = Mode.WRITE)
  public void goeBURST(@Name("project") String project, 
        @Name("dataset") String dataset, 
        @Name("lvs") long lvs, 
        @Name("inference") String inference) {
        
    InferenceService service = new InferenceService(database, log);
    service.goeBURST(project, dataset, inference, lvs);
  }
}
\end{lstlisting}

\subsection{New Service}
Listing~\ref{lst:Service} shows the common abstract class \texttt{Service}. 
Considering the goeBURST example, the \texttt{InferenceService} type is defined as a sub-class of \texttt{Service}, executing all steps required to run the goeBURST algorithm, as depicted in Listing~\ref{lst:InferenceService}.
\begin{lstlisting}[caption=Service., label=lst:Service, frame=single, basicstyle=\scriptsize]
public abstract class Service {
  public GraphDatabaseService database;
  public Log log;

  public Service(GraphDatabaseService database, Log log) {
    this.database = database;
    this.log = log;
   }
}
\end{lstlisting}
\begin{lstlisting}[caption=InferenceService., label=lst:InferenceService, frame=single, basicstyle=\scriptsize]
public class InferenceService extends Service {
/*...*/
  public void 
    goeBURST(String project, String dataset, String analysis, long lvs){
      InferenceRepository repository = new InferenceRepository(database);
      GoeBURST algorithm = new GoeBURST();
      algorithm.init(project, dataset, analysis, lvs);
      Matrix matrix;
      try (Transaction tx1 = database.beginTx()) {
        matrix = repository.read(tx1, project, dataset);
        tx1.commit();
      }
      Inference inference = algorithm.compute(matrix);
      try (Transaction tx2 = database.beginTx()) {
        repository.write(tx2, inference);
        tx2.commit();
      }
   }   
}
\end{lstlisting}
The goeBURST algorithm is implemented in \verb+GoeBURST+, which implements the helper interface \verb+Algorithm+. This interface defines the minimal set of operations that an algorithm should support.

\subsection{New Repository}
The \texttt{Repository} abstract type is provided in order to define the common read and write methods, as partially depicted in Listing~\ref{lst:Repository}.
\begin{lstlisting}[caption=Repository., label=lst:Repository, frame=single, basicstyle=\scriptsize]
public abstract class Repository<T, R> {
  /*...*/
  public abstract R read(Transaction tx, String... params) throws Exception;
  public abstract void write(Transaction tx, T param);
}
\end{lstlisting}
Then, for the goeBURST algorithm example, it is also necessary to implement the \texttt{InferenceRepository} type, sub-type of \texttt{Repository} to provide the implementation of these methods for this algorithm.

\subsection{Jar package}
If the implementation just described is added to the existing algorithms library, a new version of this library can be created by following the build instructions explained in the wiki of the project. It is also possible to define them as an extension to the algorithms library, producing a new separated jar as usual. In both cases, the new library jar must be added to the plugins directory of Neo4j along with required dependencies.
 
\section{Impact}\label{sec:evaluation}
\name is a modular framework that provides the foundations to efficiently manage, process and analyze data in the context of large phylogenetic studies.
This is accomplished by making available a secured API and by supporting new algorithms through plugins that can be executed directly on the underlying graph database.
An important feature is that new algorithms can be implemented, modified and deployed without interfering with the API and graph database functionality or availability.
As described in the project wiki, \name can be easily deployed and managed through container orchestration, with the API being horizontally scalable.

\begin{figure*}[!t]
        \centering
        \includegraphics[scale=0.67]{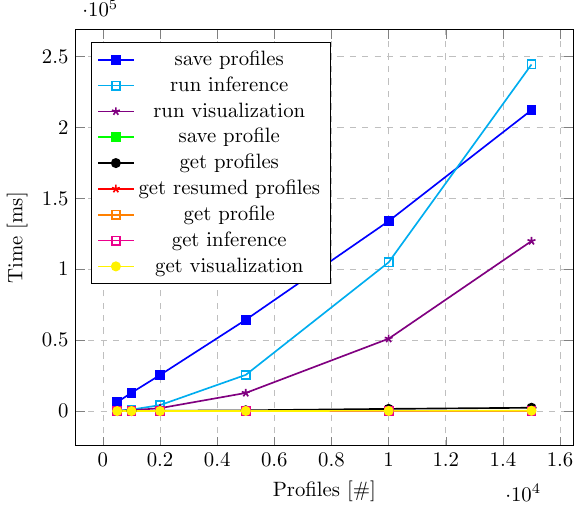}
        \includegraphics[scale=0.90]{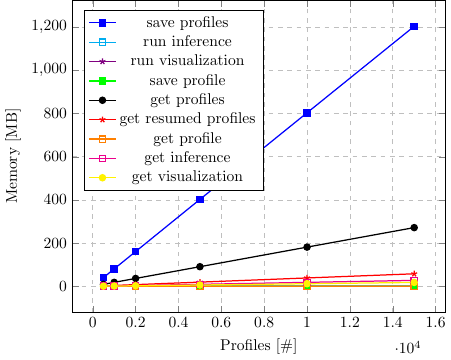}
            \caption{Plots containing the time and the memory results, in milliseconds and megabytes respectively, as new profiles are incrementally integrated for the \textit{Streptococcus pneumoniae} MLST dataset.} 
            \label{fig:plots} 
\end{figure*}

\name is currently being integrated within PHYLOViZ Online~\cite{artc:phylovizonline}. It has permitted to offload optimization tasks from the client side, and to deal with larger datasets.
Results concerning the performance of \name on real data is available in Figure~\ref{fig:plots}.
The tests were executed several times over a set of read and write operations, over the \textit{Streptococcus pneumoniae} MLST dataset~\cite{web:pubmlst}, which was specifically chosen because it is part of several published studies and also because it is publicly available, which will facilitate the interpretation and reproducibility of the results. 
Note that inference and visualization optimization operations not only include the execution of the algorithm but also the work of gathering the data and storing the results, which depends on several transactions in the underlying database. Obtained results confirm that the operations that deal with a fixed amount of data are not affected by the increasing volume of the data stored. 
Additionally, we note that execution times for the algorithms comply with their expected time complexity. This showcases that relying on a graph database to handle this type of data enables good performance and scalability.

\name allows also now to make available algorithms and methods for the dynamic updating of inferred phylogenetic trees and networks, as well as visualizations. Although previous work is known~\cite{francisco2017dynamic}, such approach is only applicable in practice if algorithms are possible to run in a setting as provided by \name. Future work comprises then the development and implementation of algorithms supporting dynamic updating of phylogentic inferences and visualizations, fundamental for large scale analyses.

\section{Conclusions}
\label{sec:conclusion}
Epidemics have become an issue of increasing importance~\cite{surveillances2020epidemiological} due to the growing exchanges of people and merchandise between countries. 
Hence, phylogenetic analyses are continuously generating huge volumes of typing and ancillary data. 
There is no doubt about the importance of such data, and phylogenetic studies, for the surveillance of infectious diseases and the understanding of pathogen population genetics and evolution. 
And the traditional way of performing phylogenetic analyses is becoming unfeasible given the amount of data generated.
The goal of this work was to develop a foundational framework for computational phylogenetic analyses that exploits graph databases with traits such as those of Neo4j.

The implementation of this framework provides a data model that is designed to represent the relationships between the several types of data and to consider multilayer networks, which enable superimposing  multiple inference results and visualizations.
This data model also contemplates versioning and soft-delete operations, allowing to keep the history of evolving datasets and analyses. 
Furthermore, the API implementation considers several other requirements, such as importing and exporting of datasets, logging, error handling and security. 
Finally, the implementations of algorithms are based on the user-defined procedures feature of Neo4j, which allows the extension of its semantic with new algorithm plugins.

\section*{Acknowledgements}
\label{}
The work reported in this article as received funding from the European Union’s Horizon
2020 research and innovation program under Grant Agreement No. 951970
(OLISSIPO project). It was also supported by national funds through Fundação para a Ciência e a Tecnologia (FCT) with reference UIDB/50021/2020.

\bibliographystyle{elsarticle-num} 
\bibliography{reference}

\end{document}